\documentclass[aps,pre,reprint,superscriptaddress,amsmath,amssymb,showpacs,floatfix,twocolumn]{revtex4-1}
\usepackage{graphicx}
\usepackage{bm}
\usepackage{color}
\usepackage{soul}
\usepackage{dcolumn}
\usepackage{multirow}
\usepackage{tikz}
\usepackage{hhline}
\usepackage{mathtools}

\definecolor{mauve}{rgb}{0.58,0,0.82}

\begin{document}
\title{ Speed Limits for Discrete-Time Markov Chains}

\author{Sangyun Lee}
\affiliation{School of Physics, Korea Institute for Advanced Study, Seoul, 02455, Korea}
\altaffiliation{Present address:
Institute for Physics, Johannes Gutenberg University Mainz, Germany.}
\affiliation{Institute for Physical Science and Technology, University of Maryland, College Park, Maryland 20742, USA}

\author{Jae Sung Lee}
\email[]{jslee@kias.re.kr}
\affiliation{School of Physics, Korea Institute for Advanced Study, Seoul, 02455, Korea}

\author{Jong-Min Park}
\email[]{jongmin.park@apctp.org}
\affiliation{Asia Pacific Center for Theoretical Physics, Pohang 37673, Korea}
\affiliation{Department of Physics, POSTECH, Pohang 37673, Korea}

\date{\today}

\begin{abstract}
As a fundamental thermodynamic principle, speed limits reveal the lower bound of entropy production (EP) required for a system to transition from a given initial state to a final state. While various speed limits have been developed for continuous-time Markov processes, their application to discrete-time Markov chains remains unexplored. In this study, we investigate the speed limits in discrete-time Markov chains, focusing on two types of EP commonly used to measure the irreversibility of a discrete-time process: time-reversed EP and time-backward EP. We find that time-reversed EP satisfies the speed limit for the continuous-time Markov processes, whereas time-backward EP does not. Additionally, for time-reversed EP, we derive practical speed limits applicable to systems driven by cyclic protocols or with unidirectional transitions, where conventional speed limits become meaningless or invalid. We show that these relations also hold for continuous-time Markov processes by taking the time-continuum limit of our results. Finally, we validate our findings through several examples: toy models, experimentally established cell-state dynamics, and theoretical computation.
\end{abstract}

\maketitle

\section{introduction}

Entropy production (EP) is a central thermodynamic quantity that reveals the irreversibility inherent in non-equilibrium processes. The mathematical formulation of EP for microscopic stochastic processes has been developed within the framework of stochastic thermodynamics~\cite{seifert2005entropy,seifert2012stochastic}, leading to the discovery of invaluable thermodynamic relations. The representative one is fluctuation theorems~\cite{jarzynski1997nonequilibrium, crooks1999entropy, seifert2005entropy,Esposito2010, Kurchan2012, Spinney2012, HyunKeun2013}, which are expressed as equalities, offering more detailed information about EP compared to the second law of thermodynamics. They represent fundamental symmetries in the distributions of thermodynamic quantities and provide experimental methods to measure the free energy difference between two states from nonequilibrium processes~\cite{jarzynski1997nonequilibrium, crooks1999entropy,collin2005verification}. 

In addition to the fluctuation theorems, numerous thermodynamic tradeoff relations have been discovered, surpassing the bound of the second law. These include the thermodynamic uncertainty relation (TUR)~\cite{Seifert2015PRL,Hasegawa2019PRL,Horowitz2020Review,Koyuk2020,Liu2020Thermodynamics,Vu2021PRL,Pal2021Thermodynamics,JSLee2021,Pal2021thermodynamic2,park2021thermodynamic}, entropic bounds~\cite{dechant2018entropic}, and fluctuation-response inequalities~\cite{DechantPNAS,ejkwonFRI}. These relations establish non-trivial lower bounds on EP in terms of generic non-equilibrium currents. They are instrumental in deriving power-efficiency tradeoff relations~\cite{Shiraish2016, pietzonka2018universal,JSLee2020, Oh2023Effects} and bounds on the turnover time scale of anomalous diffusions~\cite{hartich2021thermodynamic}. Moreover, these relations enable the inference of EP from experimental data without necessitating knowledge of the governing equations~\cite{Fakhri2019,kim2022estimating,OtsuboCommu2022,kim2022estimating,SYLee2023,Kwon2023Divergence}.

Another important finding is the discovery of thermodynamic speed limits~\cite{Shiraishi2018, Gupta2020Tighter, delvenne2021thermo, lee2022speed, Ito2020Stochastic,Vu2021PRL,Dechant2022,Van2023Thermodynamic}. It establishes a bound on EP in terms of probability distance and the dynamical activity. 
Recently, the speed limit has been used to derive the finite-time Landauer's bound~\cite{delvenne2021thermo,lee2022speed,Van2023Thermodynamic,Proesmans2} and interpret it within a unified hierarchy of tradeoff relations.~\cite{Kwon2024unified}.

Previous studies on speed limits have primarily focused on continuous-time Markov processes described by Master equations. However, discrete-time models are widely used across various research fields due to both practical and conceptual considerations. For example, the dynamics of interest may inherently evolve in discrete steps, as seen in generational updates in population~\cite{fisher1999genetical} or the evolution of binary states in computational algorithms~\cite{Gonzalo2024Thermodynamics}. Even when the underlying processes are continuous, discrete-time models are often employed to analyze empirical data collected at fixed intervals, such as data from financial time series~\cite{bischi2000nonlinear}, cell culture experiments~\cite{gupta2011stochastic}, and climate time series~\cite{Lee2024Network}.
Additionally, some coarse-grained discrete-time processes are used to infer physical quantities of the underlying continuous processes, such as a return map for the Lyapunov exponent~\cite{Packard1980Geometry} and a milestoned process for entropy production~\cite{blom2024milestoning}.
Finally, some rule-based discrete-time models offer valuable theoretical insights with enhanced analytical tractability, contributing to the understanding of the routes to chaos~\cite{may1976simple}, self-organized criticality~\cite{bak1988self}, and nonequilibrium phase transitions~\cite{achlioptas2009explosive}.

Despite the importance of discrete-time Markov processes, relatively few studies have addressed EP and thermodynamic relations in such systems.
Some papers have defined EP in discrete-time Markov chains by relating it to the degree of irreversibility~\cite{gaspard2004time,ge2006reversibility} and the violation of detailed balance, while others have addressed discrete-time EP as part of their primary goal of extending the TUR to discrete-time processes~\cite{shiraishi2017finite, proesmans2017discrete,Hasegawa2019PRL,timpanaro2019thermodynamic,Liu2020Thermodynamics}.
In particular, the violation of the TUR in discrete-time processes was first demonstrated in~\cite{shiraishi2017finite}, and explicit formulations were derived in subsequent studies~\cite{proesmans2017discrete,Hasegawa2019PRL,timpanaro2019thermodynamic,Liu2020Thermodynamics}.
However, the extensions of other thermodynamic relations, in particular speed limits, to discrete-time processes remain underexplored.

In addition, previous studies on speed limits have focused on systems without unidirectional transitions.
However, systems with unidirectional transitions are often used to describe dynamics where reverse transitions are either extremely rare or missing due to incomplete experimental data~\cite{gupta2011stochastic,baiesi2024effective}.
Physically relevant examples include totally asymmetric exclusion processes~\cite{derrida1993exact, saha2016entropy, Soh2018Jamming}, directed percolation~\cite{takeuchi2007directed}, spontaneous emission~\cite{rahav2014integral}, opinion dynamics~\cite{holley1975ergodic}, stochastic resetting~\cite{Evans2011Diffusion, fuchs2016stochastic, tal2020experimental, chen2022first, mori2023entropy}, and theoretical computation~\cite{Gonzalo2024Thermodynamics}, as well as various biological~\cite{gupta2011stochastic,dharmaraja2015continuous,roldan2016stochastic,bressloff2020directed} and chemical processes~\cite{michaelis1913kinetik,ting2009kinetic,johnson2011original}. While the TUR has been extended to systems with unidirectional transitions~\cite{Pal2021Thermodynamics}, a corresponding extension for speed limits remains unexplored.
In this context, it is intriguing and important to investigate the effects of time discretization and unidirectional transitions on speed limits.

Here, we study speed limits for discrete-time Markov chains. In the literature on discrete-time processes, at least two EP formulations can be found: time-reversed EP and time-backward EP~\cite{Liu2020Thermodynamics,Lee2018Derivation,gaspard2004time,igarashi2022entropy,proesmans2017discrete, touchette2009large}, both of which converge to the same value in the continuous-time limit.
We investigate speed limits for these EPs and verify that the speed limit derived for continuous-time Markov processes still holds for time-reversed EP, while it is violated for time-backward EP. We quantify the degree of this violation using the Kullback-Leibler divergence between consecutive probability distributions. This indicates that the bound of the speed limit depends on the choice of EP formulation and that modification from the conventional form is required for time-backward EP. We also find the saturation conditions of the speed limits for time-reversed EP.

We extend the thermodynamic speed limit to two types of processes where the conventional speed limit fails or becomes trivial: (i) systems that return to their initial distribution~\cite{Shiraishi2018,lee2022speed} and (ii) systems with unidirectional transitions~\cite{Gonzalo2024Thermodynamics, Busiello2020Entropy, rahav2014integral}.
In the first case, which is often used to describe thermal devices operating in a cyclic steady state~\cite{Feldmann2003Quantum, Park2019Quantum, Feldmann2000Performance, Lee2020Finite, Lee2021Quantumness, chun2025power}, the probabilistic distance between the identical initial and final states is zero, causing the conventional speed limit bound to vanish and become meaningless.
To address this limitation, we derive tighter bounds expressed in terms of either the sum of distances between consecutive states or the distance between two states at any intermediate time.

This article is organized as follows. In Sec.~\ref{sec:Discretetimespeedlimit}, we review the two formulations of EP for a discrete-time Markov chain. In Sec.~\ref{sec:speed_limit_derivation}, we present the derivation of speed limits for these EPs in discrete-time Markov chains. Specifically, we derive the stepwise speed limit, which requires the full information of probability distributions at all time steps, and the interval speed limit, which necessitates only partial information on probability distributions at limited time steps. The saturation condition for the speed limits is also discussed. In Sec.~\ref{sec:uni-SL}, we extend our result to a Markov chain with unidirectional transitions.
In Sec.~\ref{sec:Numerics}, we numerically verify our results through three examples. Finally, we conclude in Sec.~\ref{sec:disscussion}.

\section{Entropy productions}
\label{sec:Discretetimespeedlimit}

Before discussing the discrete-time speed limits, we first revisit the formulations of EP for a discrete-time Markov chain. A few references have explored the EP of a discrete-time Markov chain, with at least two types of EP being studied~\cite{Liu2020Thermodynamics, Lee2018Derivation, gaspard2004time, igarashi2022entropy, proesmans2017discrete, 
 touchette2009large}. To explain these EPs, we introduce our notations for the discrete-time Markov chain.

We denote the time at which the $i$-th transition occurs as $t_i$ for $i=0,1,...,N$, with a time gap $\Delta t = t_{i+1}-t_i$, and the probability of being in state $n$ at time $t_i$ by $P_n(t_i)$.
The Markov chain is characterized by the transition probability $M_{n,m} (t_i)$, denoting the conditional probability of transition from the state $m$ at $t_{i}$ to $n$ at $t_{i+1}$, $M_{n,m} (t) \equiv P(n,t_{i+1}| m,t_i) $.
We assume that every transition has a corresponding reverse transition, i.e., $M_{n,m} (t) \neq 0$ if $M_{m,n} (t) \neq 0$.
The case involving unidirectional transitions will be discussed in Sec.~\ref{sec:uni-SL}.

In terms of $P_n(t_i)$ and $M_{n,m}(t_i)$, the probability of being in state $n$ at $t_{i+1}$ is given by the following equation:
\begin{equation} \label{eq:discrete_master_eq}
    P_n(t_{i+1}) = \sum_m M_{n,m}(t_i) P_m(t_i).
\end{equation}
Using the normalization condition of the transition probability, $\sum_n M_{n,m}(t_i) = 1$, this equation can be rearranged as
\begin{align}\label{eq:alt_master_eq}
    P_n&(t_{i+1}) - P_n(t_i) \nonumber \\
    &= \sum_{m(\neq n)}
    \left[ 
    M_{n,m} (t_i) P_m(t_i)
    - M_{m,n} (t_i) P_n(t_i)
    \right ]
.\end{align} Equation~\eqref{eq:alt_master_eq} will be utilized to derive our discrete-time speed limits.
This equation is analogous to the following master equation for a continuous-time Markov process: 
\begin{align}
    \dot P_n(t) = \sum_{m(\neq n)} [R_{n,m}(t)P_m(t) -R_{m,n}(t)P_n(t)],
\label{eq:conti_master}
\end{align} 
where $R_{n,m}(t)$ is the transition rate at time $t$. By dividing Eq.~\eqref{eq:alt_master_eq} by $\Delta t$ and taking the continuum limit, one can derive Eq.~\eqref{eq:conti_master} from Eq.~\eqref{eq:alt_master_eq} when $M_{n,m}(t_i) = \delta_{n,m} + R_{n,m} (t_i) \Delta t + O(\Delta t^2)$. The results of this paper can also be extended to deduce continuous-time results.

We note that, in general, time may not have a relevant physical meaning in discrete-time processes. For example, in computational algorithms, state transitions occur after each logical operation, without requiring a fixed time interval. In such cases, the discrete-time $t_i$ merely denote the index corresponding to the number of operations. For simplicity, we focus on cases where the time interval is fixed, while our results can be straightforwardly extended to general scenarios.

For discrete-time processes, two definitions of EP have been used in literature~\cite{igarashi2022entropy}: (i) time-reversed EP $\tilde{\Sigma}$~\cite{ gaspard2004time,touchette2009large,proesmans2017discrete} and (ii) time-backward EP $\Sigma$~\cite{Lee2018Derivation, Liu2020Thermodynamics}, which are given by
\begin{align}
    \Delta \tilde{\Sigma} (t_i)&\equiv 
     \sum_{n,m} M_{n,m} (t_i) P_m(t_i) \ln \frac{M_{n,m} (t_i) P_m(t_i)}{M_{m,n} (t_i) P_n(t_i)},    \label{def:Delta_rev_EP} \\
    \Delta \Sigma (t_i) &\equiv \sum_{n,m} M_{n,m} (t_i) P_m(t_i) \ln \frac{M_{n,m} (t_i) P_m(t_i)}{M_{m,n} (t_i) P_n(t_{i+1})} .
    \label{def:Delta_back_EP}
 \end{align}
Notably, their difference lies only in the time argument of $P_n(t)$.
Thus, the time-reversed and time-backward EPs are identical when the system is in a steady-state  $P_n(t) = P_n$.

By using the log inequality $\ln x+1-x\le 0$ for $x > 0$, one can show that both EPs are always non-negative regardless of the time interval. In addition, $\Delta \tilde{\Sigma} \geq \Delta \Sigma$ can be proved by using the same log inequality as follows:
\begin{equation}\label{eq:relationbtwtimeandback}
    \Delta \Sigma'(t_i) \equiv 
    \Delta \tilde{\Sigma}(t_i) - \Delta \Sigma(t_i)
    = D[ \mathbf{P} (t_{i+1}) \parallel \mathbf{P} (t_i)] \geq 0,
\end{equation}
where $D[ \mathbf{p} \parallel \mathbf{q}] = \sum_n p_n \ln(p_n/q_n)$ denotes the Kullack-Leibler divergence of a probability $p_{n}$ with respect to a probability $q_{n}$.
In the continuous-time limit, their difference, $\Delta \Sigma'$, vanishes as the order of the square of the time interval. Thus, both EPs become identical in that limit.

Due to the difference in $P_n(t)$, the time-reversed and time-backward EP exhibit several significant differences. For instance, some properties of the EP rate in continuous-time processes are shared not by both, but solely by time-reversed EP.
Their differences are explained in Appendix~\ref{app:differences_EPs}.

\section{Discrete-time speed limits}
\label{sec:speed_limit_derivation}

\subsection{Stepwise speed limit}

In this section, we describe the process of deriving speed limits for discrete-time systems, focusing initially on time-reversed EP~$\tilde{\Sigma}$. The speed limit can be derived through the method used in previous works~\cite{delvenne2021thermo,lee2022speed}, since the time-reversed EP shares the same mathematical structure as the EP rate for continuous-time processes. Thus, we can express $\Delta \tilde{\Sigma} (t_i)$ as the following Kullback-Leibler divergence form:
\begin{align}\label{eq:rev_EP_KLDform}
    \Delta \tilde{\Sigma} (t_i)
    &= A(t_i) \sum_{n,m}
    Q_{n,m} (t_i)
    \ln \frac{Q_{n,m} (t_i)}{Q_{m,n} (t_i) }\nonumber\\
    &= A(t_i) D [ \mathbf{Q} (t_i) \parallel \mathbf{Q}^\textsf{T} (t_i) ],
\end{align}
where $Q_{n,m}(t_i)$ is a joint probability defined as $Q_{n,m} (t_i) \equiv M_{n,m} (t_i) P_m (t_i) / A(t_i)$ with normalization condition $\sum_{n\neq m}Q_{n,m}(t_i) =1$, $A(t_i) \equiv \sum_{n\neq m} M_{n,m} (t_i) P_m (t_i)$ is the dynamical activity, and $\textsf{T}$ denotes the transpose of a matrix, i.e.,
$Q^\textsf{T}_{n,m} (t_i) = Q_{m,n} (t_i)$. The Kullback--Leibler divergence between two matrices is defined as
\begin{align}
    D[\mathbf A \parallel \mathbf B] \equiv \sum_{n,m} A_{n,m}
    \ln \frac{A_{n,m}}{B_{n,m}}
.\end{align}

Several inequalities establish the relation between the Kullback-Leibler divergence and the total variation distance, which is a measure of the distance between two distributions $\mathbf{p}$ and $\mathbf{q}$, defined as
\begin{align} \label{def:TVD}
    d_\textrm{TV} [\mathbf{p}, \mathbf{q}] \equiv \frac{1}{2} \sum_n |p_n - q_n|.
\end{align}
For example, Pinsker's inequality~\cite{Pinsker60} and Gilardoni's inequality~\cite{Gilardoni08} provide the bounds expressed as
$D[\mathbf{p} \parallel \mathbf{q}] \geq 2 (d_\textrm{TV} [\mathbf{p}, \mathbf{q}])^2$
and
$D[\mathbf{p} \parallel \mathbf{q}] \geq h_\textrm{G} \left( d_\textrm{TV} [\mathbf{p}, \mathbf{q}] \right)$
with $h_\textrm{G}(x)=\ln [(1+x)^{-1+x}/(1-x)]$, respectively.
Which of these two inequalities is tighter depends on the value of $d_\textrm{TV} [\mathbf{p}, \mathbf{q}]$~\cite{lee2022speed}.
When the Kullback-Leibler divergence is symmetric
($D[\bm p|| \bm q] = D[\bm q|| \bm p ]$), a tighter bound can be derived~\cite{gilardoni2006minimum,Gilardoni2008,falasco2022beyond}.
This bound can be obtained from the bound for the symmetrized Kullback–Leibler divergence, also known as the Jeffreys divergence $D_{\rm J}[\mathbf{p},\mathbf{q}]$ (See Appendix.~\ref{app:derivation_symGilardoni} for its derivation):
\begin{equation}\label{ineq:Jeffreys_vs_dTV}
D_{\rm J}[\mathbf{p},\mathbf{q}] \equiv \frac{D[\mathbf{p} \parallel \mathbf{q}] + D[\mathbf{q} \parallel \mathbf{p}]}{2}
 \geq h \left( d_\textrm{TV} [\mathbf{p}, \mathbf{q}] \right)
\end{equation}
where
\begin{align}
h(x) \equiv x \ln \frac{1+x}{1-x}.
\label{eq:hsym}
\end{align}
Thus, if Kullback-Leibler divergence is symmetric, $D_\textrm{J} [\mathbf{p}, \mathbf{q}] =  D[\mathbf{p} \parallel \mathbf{q}]$, we obtain~\cite{gilardoni2006minimum,Gilardoni2008,lee2022speed}
\begin{equation}\label{ineq:KL_vs_dTV}
D[\mathbf{p} \parallel \mathbf{q}] \geq h \left( d_\textrm{TV} [\mathbf{p}, \mathbf{q}] \right).
\end{equation}
We note that $h(x)$ is convex and a monotonically increasing function for $x\geq0$. 
Since
\begin{align}
    D[\mathbf{Q}^\textrm{T} (t_i) \parallel \mathbf{Q} (t_i)]
    &= \sum_{n,m} Q_{m,n} (t_i) \ln \frac{Q_{m,n} (t_i)}{Q_{n,m} (t_i)}
    \nonumber \\
    &= \sum_{n,m} Q_{n,m} (t_i) \ln \frac{Q_{n,m} (t_i)}{Q_{m,n} (t_i)}
    \nonumber \\
    & = D[\mathbf{Q} (t_i) \parallel \mathbf{Q}^\textrm{T} (t_i)],
\end{align}
the Kullback-Leibler divergence in Eq.~\eqref{eq:rev_EP_KLDform} is symmetric.
Consequently, we can apply Eq.~\eqref{ineq:KL_vs_dTV} to Eq.~\eqref{eq:rev_EP_KLDform}, which yields
\begin{equation}\label{ineq:rev_EP_KLineq}
    \Delta \tilde \Sigma (t_i) 
    \geq A(t_i) h \left ( d_{\rm TV} [ {\mathbf Q}(t_i),\mathbf{Q}^\textrm{T}(t_i) ] \right ) 
.\end{equation}
As the next step, by employing the following triangle inequality 
\begin{align}\label{ineq:triangle_current}
    \sum_{m (\neq n)} &\big | M_{n,m} (t_i) P_m (t_i) - M_{m,n} (t_i) P_n (t_i) \big | \nonumber \\
    &\geq \left | \sum_{m (\neq n)} \left[ M_{n,m} (t_i) P_m (t_i) - M_{m,n} (t_i) P_n (t_i) \right] \right |
\end{align}
and Eq.~\eqref{eq:alt_master_eq}, we can derive the lower bound of the total variation distance for $\mathbf Q (t_i)$ as
\begin{align}\label{ineq:dTV_Q_dTV_P}
    d_{\rm TV} [ {\mathbf Q}(t_i),\mathbf{Q}^\textsf{T}(t_i) ]
    &= \sum_n  \frac{\sum_{m (\neq n)} \left| Q_{n,m} (t_i)  - Q_{m,n}(t_i) \right |}{2} \nonumber \\
    &\geq \sum_n  \frac{\left| P_n (t_{i+1}) - P_n(t_i) \right |}{2 A(t_i)} = \frac{d_{i+1,i}}{A(t_i)}
\end{align} 
where $d_{i+1,i} \equiv d_{\rm TV} [\mathbf{P} (t_{i+1}), \mathbf{P} (t_i)]$.
Plugging Eq.~\eqref{ineq:dTV_Q_dTV_P} into Eq.~\eqref{ineq:rev_EP_KLineq} under the monotonically increasing property of $h(x)$ and summing over all $i$ yields
\begin{align}\label{ineq:StepWise_SpeedLimit_rev_EP}
    \tilde \Sigma \equiv \sum_{i=0}^{N-1} \Delta \tilde \Sigma(t_i) 
    \geq \sum_{i=0}^{N-1}A(t_i) h\left ( \frac{d_{i+1,i}}{A(t_i)} \right ).
\end{align}
We call equation~\eqref{ineq:StepWise_SpeedLimit_rev_EP} stepwise speed limit, as the lower bound of the entire EP is estimated by summing all time-step bounds. 

By treating the bound as the average of a convex function $h(x)$ over a distribution $A (t_i)/A_\textrm{tot}$ with $A_\textrm{tot} \equiv \sum_{i=0}^{N-1} A (t_i)$, one can apply the Jensen's inequality as follows:
\begin{equation}\label{ineq:Jensen_dTV}
    \sum_{i=0}^{N-1} \frac{A(t_i)}{A_\text{tot}}
    h \left (  \frac{d_{i+1,i}}{A(t_i)} \right )
    \geq h \left ( \sum_{i=0}^{N-1} \frac{d_{i+1,i}}{A_\textrm{tot}} \right ).
\end{equation}
Applying  Eq.~\eqref{ineq:Jensen_dTV} into Eq.~\eqref{ineq:StepWise_SpeedLimit_rev_EP}, 
we finally arrive at 
\begin{align}
    \tilde \Sigma \geq& A_{\rm tot } h\left ( \sum_{i=0}^{N-1}\frac{d_{i+1,i}}{A_{\rm tot}  } \right )
\label{eq:speedlimitfirstform}.
\end{align}
This expression is applicable when $\mathbf{P}(t_0) = \mathbf{P}(t_N) $  and yields a nontrivial bound for this case. We note that $A_\textrm{tot}$ can be computed by averaging the number of transitions between all different states without the knowledge of $\mathbf{P}(t)$. On a related note, a similar relation in terms of the Wasserstein distance for continuous-time processes, analogous to Eq.\eqref{eq:speedlimitfirstform}, was reported in Ref.\cite{Van2024geometric}.

\subsection{Interval speed limit }

To evaluate the bound of the stepwise speed limit~\eqref{eq:speedlimitfirstform}, full information on $\mathbf{P}(t)$ at all time steps is required. However, acquiring all the information on the temporal evolution of $\mathbf{P}(t)$ is not always possible. There may be cases where partial information on $\mathbf{P}(t)$ at limited time steps is available. Hence, it would be useful to find alternative bounds utilizing less information on $\mathbf{P}(t)$. Here, we present a speed limit using the knowledge of probability distributions only at two time points. To this end, we employ the triangle inequality for the total variation distance as follows:
\begin{equation}\label{ineq:triangle_dTV}
    \sum_{i=0}^{N-1} d_{i+1,i}
    \geq d_\textrm{TV} [\mathbf{P}(t_{\rm F}),\mathbf{P}(t_{\rm I})]
\end{equation}
for any intermediate times $t_{\rm I}$ and $t_{\rm F}$ within $[t_0,t_N]$. Then, from the monotonically increasing property of $h(x)$, a simplified bound can be obtained by
\begin{equation}\label{ineq:TwoPoint_SpeedLimit_rev_EP}
    \tilde{\Sigma} \geq A_\textrm{tot}
    h \left( \frac{d_\textrm{TV}[\mathbf{P}(t_{\rm F}),\mathbf{P}(t_{\rm I})]}{A_\textrm{tot}} \right).
\end{equation} 
We call Eq.~\eqref{ineq:TwoPoint_SpeedLimit_rev_EP} interval speed limit. By using the relation $h(x) = 2x\tanh^{-1}x$, one can rearrange Eq.~\eqref{ineq:TwoPoint_SpeedLimit_rev_EP} in the following conventional form of a speed limit:
\begin{equation}
    t_{N} \geq \frac{d_\textrm{TV} [ \mathbf{P}(t_{\rm F}), \mathbf{P}(t_{\rm I}) ]}{\bar{A} \tanh \left( \frac{\tilde{\Sigma
    }}{2 d_\textrm{TV} [\mathbf{P} (t_{\rm F}), \mathbf{P}(t_{\rm I})]} \right)},
\label{eq:speedlimit_conv}\end{equation}
where $\bar{A}\equiv A_\textrm{tot}/t_{N}$. When we set $t_{\rm I} = t_0$ and $t_{\rm F} = t_N$, Eq.~\eqref{eq:speedlimit_conv} has the same mathematical form as the continuous-time speed limit~\cite{delvenne2021thermo, JSLee2021}.
In general cases where the physical time is irrelevant,
a bound for the number of steps can be expressed as
\begin{align}
        N \geq \frac{d_\textrm{TV} [ \mathbf{P}(t_{\rm F}), \mathbf{P}(t_{\rm I}) ]}{\bar{A}_N \tanh \left( \frac{\tilde{\Sigma
    }}{2 d_\textrm{TV} [\mathbf{P} (t_{\rm F}), \mathbf{P}(t_{\rm I})]} \right)}
    \label{eq:step_bound}
\end{align}
with the average total activity $\bar{A}_N \equiv A_{\rm tot}/N$ per unit transition.

It is worth comparing the EP bound derived from our results with the TUR for discrete-time Markov chains. The EP lower bound of the discrete-time TUR is exponentially suppressed~\cite{proesmans2017discrete} or proportional to the minimum staying probability~\cite{Liu2020Thermodynamics}. Moreover, the existing discrete-time TUR  applies only to systems without time-dependent driving, whereas our speed limit has no such constraint. This highlights the potential utility of our result for estimating EP in discrete-time processes with arbitrary driving protocols. 

In addition, the interval speed limit  can yield a nontrivial bound for cyclic processes, whereas the conventional speed limit becomes trivial due to the complete cycle condition $\mathbf{P}(t_0) = \mathbf{P}(t_N)$ in a cyclic process. We present a concrete application of the interval speed limit  on a cyclic steady state in Sec.~\ref{sec:Numerics}.

Using Eq.~\eqref{eq:relationbtwtimeandback},
one can show that
\begin{align}
    \Sigma = \sum_{i=0}^{N-1} \left(\Delta \tilde{\Sigma}(t_i) - \Delta \Sigma' (t_i) \right) = \tilde{\Sigma} - \sum_{i=0}^{N-1} \Delta \Sigma' (t_i) .
\end{align}
It yields the speed limit for the time-backward EP given by
\begin{equation}\label{ineq:TwoPoint_SpeedLimit_back_EP}
    \Sigma \geq A_\textrm{tot}
    h \left( \frac{d_\textrm{TV}[\mathbf{P}(t_{\rm F}),\mathbf{P}(t_{\rm I})]}{A_\textrm{tot}} \right) - \sum_{i=0}^{N-1} 
    \Delta \Sigma'(t_i).
\end{equation}
Evaluating the last term on the right-hand side requires information about distributions at all time steps. Furthermore, the lower bound of the time-backward EP is smaller than that of the time-reversed EP, indicating that the time-backward EP violates the conventional speed limit. In Sec.~\ref{subsec:example1}, we will demonstrate an example of this violation in a simple two-state system.

\subsection{Saturation condition} \label{sec:saturation}
We now investigate the condition for each speed limit being saturated. They saturate when the equality conditions for all inequalities used to derive each speed limit are satisfied. In this regard, inequalities~\eqref{ineq:KL_vs_dTV} and \eqref{ineq:triangle_current} are used to derive the stepwise speed limit, and  additional inequalities~\eqref{ineq:Jensen_dTV} and \eqref{ineq:triangle_dTV} are applied for the interval speed limit . First, the equality condition of Eq.~\eqref{ineq:triangle_current} is satisfied when the system has only two states. Interestingly, for a two-state system, Eq.~\eqref{ineq:KL_vs_dTV} also saturates regardless of the protocol $M_{n,m} (t)$. Therefore, the stepwise speed limit always saturates for any two-state system.
The equal sign of Eq.~\eqref{ineq:Jensen_dTV} is attainable when
the following condition is satisfied: 
\begin{equation}
    \frac{d_{i+1,i}}{A (t_i)}
    =
    \frac{d_{j+1,j}}{A (t_j)}
\end{equation}
for any $i \neq j$. Finding the general saturation condition for inequality~\eqref{ineq:triangle_dTV} is difficult. Nonetheless, the equality of Eq.~\eqref{ineq:triangle_dTV} is trivially satisfied when the process ends after only a single time step, i.e., $N = 1$, as the interval speed limit and stepwise speed limit become identical when $N = 1$. 

In short, regardless of the protocol $M_{n,m} (t_i)$, the equality of the stepwise speed limit can be achieved for two-state systems with any number of time steps, and the interval speed limit saturates in one-time step processes of a system with only two states.

\section{Speed limit with unidirectional transitions}
\label{sec:uni-SL}
We now extend our results to systems that involve both bidirectional and unidirectional transitions. A bidirectional transition refers to a transition where the corresponding backward transition exists, while a unidirectional transition refers to a transition where the backward transition is absent.
We use a superscript $\alpha = {\rm u}$ or ${\rm b}$ to represent the transition probability associated with the unidirectional or bidirectional transition, respectively. That is, $M^{({\rm u})}_{n,m} (t_i)$ and $M^{({\rm b})}_{n,m} (t_i)$ denote that the transition from $m$ state to $n$ state is unidirectional and bidirectional, respectively. Therefore, if $M^{({\rm u})}_{n,m} (t_i)$ is nonzero, $M^{({\rm u})}_{m,n} (t_i)$ must be $0$. Conversely, if $M^{({\rm b})}_{n,m} (t_i)$ is nonzero, $M^{({\rm b})}_{m,n} (t_i)$ must also be nonzero. 
With this notation, the equation of motion can be written as
\begin{equation}
    P_n (t_{i+1}) = \sum_{\alpha = {\rm u},{\rm b}} {\sum_{m }}^{(\alpha)}   
    M^{(\alpha)}_{n,m} (t_i) P_m(t_i),
\end{equation}
where $\sum_{m }^{(\alpha)}$ denotes summation over restricted $m$ values such that the transition between $m$ and $n$ belongs to $\alpha$. 

Defining EP for a system with unidirectional transitions is a nontrivial problem. Thus, several approaches have been suggested to extend the results of EP for Markovian systems with only bidirectional transitions to systems with unidirectional transitions~\cite{rahav2014integral, fuchs2016stochastic, Gupta2020Tighter, Busiello2020Entropy, mori2023entropy, Gonzalo2024Thermodynamics}.
This difficulty arises because directly applying the EP definition in Eqs.~\eqref{def:Delta_rev_EP} or \eqref{def:Delta_back_EP} to systems with unidirectional transitions leads to divergences.
A similar problem also arises in systems with absolutely irreversible processes~\cite{Murashita2014Nonequilibrium,hoang2016scaling}.
One possible resolution is to define EP solely based on bidirectional transitions.
This definition has been adopted in studies of fluctuation theorems for systems with unidirectional transitions~\cite{rahav2014integral}, as well as in other studies on continuous-time Markov processes with unidirectional transitions~\cite{fuchs2016stochastic, Busiello2020Entropy,Pal2021Thermodynamics,Gonzalo2024Thermodynamics}.

This definition of entropy production can be obtained by decomposing the system entropy change as
\begin{align}
    \Delta S_{\rm sys} (t_i)=
    \Delta S^{\rm (b)}_{\rm sys} (t_i)+ \Delta S^{\rm (u)}_{\rm sys} (t_i),
\end{align}
where
\begin{equation}
    \Delta S^{(\alpha)}_{\rm sys} (t_i) \equiv {\sum_{n,m}}^{(\alpha)} M_{n,m}^{(\alpha)}(t_i)P_m(t_i)\ln{\frac{P_m(t_i)}{P_n(t_{i+1})}}
\end{equation}
represents the entropy change associated with transition type $\alpha$.
The first term, corresponding to the bidirectional transitions, can be further decomposed as~\cite{fuchs2016stochastic, Gupta2020Tighter, Busiello2020Entropy}:
\begin{align}
    \Delta S^{\rm (b)}_{\rm sys} (t_i)=
    &\Delta \Sigma^{({\rm b})} (t_i)
    -{\sum_{n,m}}^{({\rm b})} M_{n,m}^{\rm (b)}(t_i)P_m(t_{i})\ln{\frac{M^{({\rm b})}_{n,m}(t_i)}{M^{({\rm b})}_{m,n}(t_i)}},
\end{align}
where the time-backward EP for bidirectional transitions is defined as
\begin{align}
    \Delta \Sigma^{({\rm b})} (t_i) \equiv
    {\sum_{n,m}}^{({\rm b})} M_{n,m}^{({\rm b})} (t_i) P_m(t_i) \ln \frac{M_{n,m}^{({\rm b})} (t_i) P_m(t_i)}{M_{m,n}^{({\rm b})} (t_i) P_n(t_{i+1})}.
    \label{eq:time_backward_bi}
\end{align}
This decomposition shows that the modified EP, $\Delta \Sigma^{\rm (b)}$, can be interpreted as the entropy production arising solely from the bidirectional transitions.
Similarly, the time-reversed EP only for bidirectional transitions can be defined as
\begin{align}
    \Delta \tilde{\Sigma}^{({\rm b})} (t_i) \equiv& {\sum_{n,m}}^{({\rm b})} M_{n,m}^{({\rm b})} (t_i) P_m(t_i) \ln \frac{M_{n,m}^{({\rm b})} (t_i) P_m(t_i)}{M_{m,n}^{({\rm b})} (t_i) P_n(t_i)}, 
    \label{eq:time_reversed_bi}
\end{align}
and, in terms of the time-reversed EP, the system entropy change is decomposed as 
\begin{align}
    \Delta S_{\rm sys} (t_i)
    =& \Delta \tilde{\Sigma}^{({\rm b})} (t_i) + {\sum_{n,m}}^{({\rm u})} M_{n,m}^{\rm (u)}(t_i)P_m(t_i)\ln{\frac{P_m(t_i)}{P_n(t_{i+1})}}\nonumber \\
    &-{\sum_{n,m}}^{({\rm b})} M_{n,m}^{\rm (b)}(t_i)P_m(t_{i})\ln{\frac{M^{({\rm b})}_{n,m}(t_i)P_m(t_i)}{M^{({\rm b})}_{m,n}(t_i)P_n(t_{i})}}
.\end{align}

The physical interpretation of EPs defined in Eqs.~\eqref{eq:time_backward_bi} and \eqref{eq:time_reversed_bi} depends on the specific context in which the unidirectional transitions are introduced.
For instance, in systems with stochastic resetting, this EP corresponds to the EP generated during diffusion between successive resetting events~\cite{fuchs2016stochastic}. In general, such EP quantifies the degree of irreversibility arising solely from bidirectional transitions.
In the following, we demonstrate that using this definition allows us to derive a speed limit. 

Two types of dynamical activities can be defined as 
\begin{equation}
    A^{(\alpha)} (t_i) \equiv {\sum_{m\neq n}}^{(\alpha)}
    M^{(\alpha)}_{n,m} (t_i) P_m (t_i)
\end{equation}
for $\alpha \in \{\rm u,b\}$.
Note that the dynamical activity for unidirectional transition can be written as
\begin{equation}\label{eq:activity_u}
    A^{({\rm u})} (t_i) = \frac{1}{2} {\sum_{m\neq n}}^{(\rm u)} |M^{({\rm u})}_{n,m} P_m(t_i) - M^{({\rm u})}_{m,n} P_n(t_i)|.
\end{equation}
The triangle inequality leads to
\begin{align} \label{eq:uni_triangle}
    &\sum_{\alpha = {\rm u, b}} {\sum_{m\neq n}}^{({\alpha})} |M^{(\alpha)}_{n,m} P_m(t_i) - M^{(\alpha)}_{m,n} P_n(t_i)| \nonumber \\
    &\geq \sum_n \left| \sum_{\alpha = {\rm u, b}} {\sum_{m(\neq n)}}^{({\alpha})} [M^{(\alpha)}_{n,m} P_m(t_i) - M^{(\alpha)}_{m,n} P_n(t_i) ]\right| \nonumber \\
    & = \sum_n | P_n(t_{i+1}) - P_n(t_i)| = 2d_{i+1,i}.
\end{align}
Using the relation
\begin{align}
    &\sum_{\alpha = {\rm u, b}} {\sum_{m\neq n}}^{({\alpha})} |M^{(\alpha)}_{n,m} P_m(t_i) - M^{(\alpha)}_{m,n} P_n(t_i)|  \nonumber \\
    &= {\sum_{m\neq n}}^{({\rm b})} |M^{({\rm b})}_{n,m} P_m(t_i) - M^{({\rm b})}_{m,n} P_n(t_i)|
    + 2 A^{({\rm u})} (t_i),
\end{align}
Eq.~\eqref{eq:uni_triangle} is reexpressed as
\begin{equation}
    \frac{A^{({\rm b})}(t_i)}{2}{\sum_{m\neq n}}^{(\rm b)} |Q^{({\rm b})}_{n,m}(t_i) - Q^{({\rm b})}_{m,n}(t_i) |
    \geq d_{i+1,i} - A^{(u)} (t_i),
\end{equation}
where $Q^{({\rm b})}_{n,m}(t_i) \equiv M^{({\rm b})}_{n,m}(t_i) P_m (t_i)/A^{({\rm b})}(t_i)$.  Along with this inequality, by following a similar derivation procedure as presented in Eqs.~\eqref{eq:rev_EP_KLDform}, \eqref{ineq:KL_vs_dTV}, and \eqref{ineq:StepWise_SpeedLimit_rev_EP}, we arrive at the stepwise speed limit for a system including unidirectional transitions as follows:
\begin{align} \label{eq:stepwise_sl_uni}
    \tilde \Sigma^{({\rm b})}
    \geq \sum_{i=0}^{N-1}A^{({\rm b})}(t_i) h\left( \frac{d_{i+1,i} - A^{({\rm u})} (t_i)} {A^{({\rm b})}(t_i)} \right). 
\end{align}
Note that Eq.~\eqref{eq:stepwise_sl_uni} is valid only when $d_{i+1,i} - A^{(u)} (t_i) \geq 0$ for all $t_i$, since $h(x)$ is a monotonically increasing function only for $x \geq 0$.

\begin{figure}[t]
\includegraphics[width=0.48\textwidth]{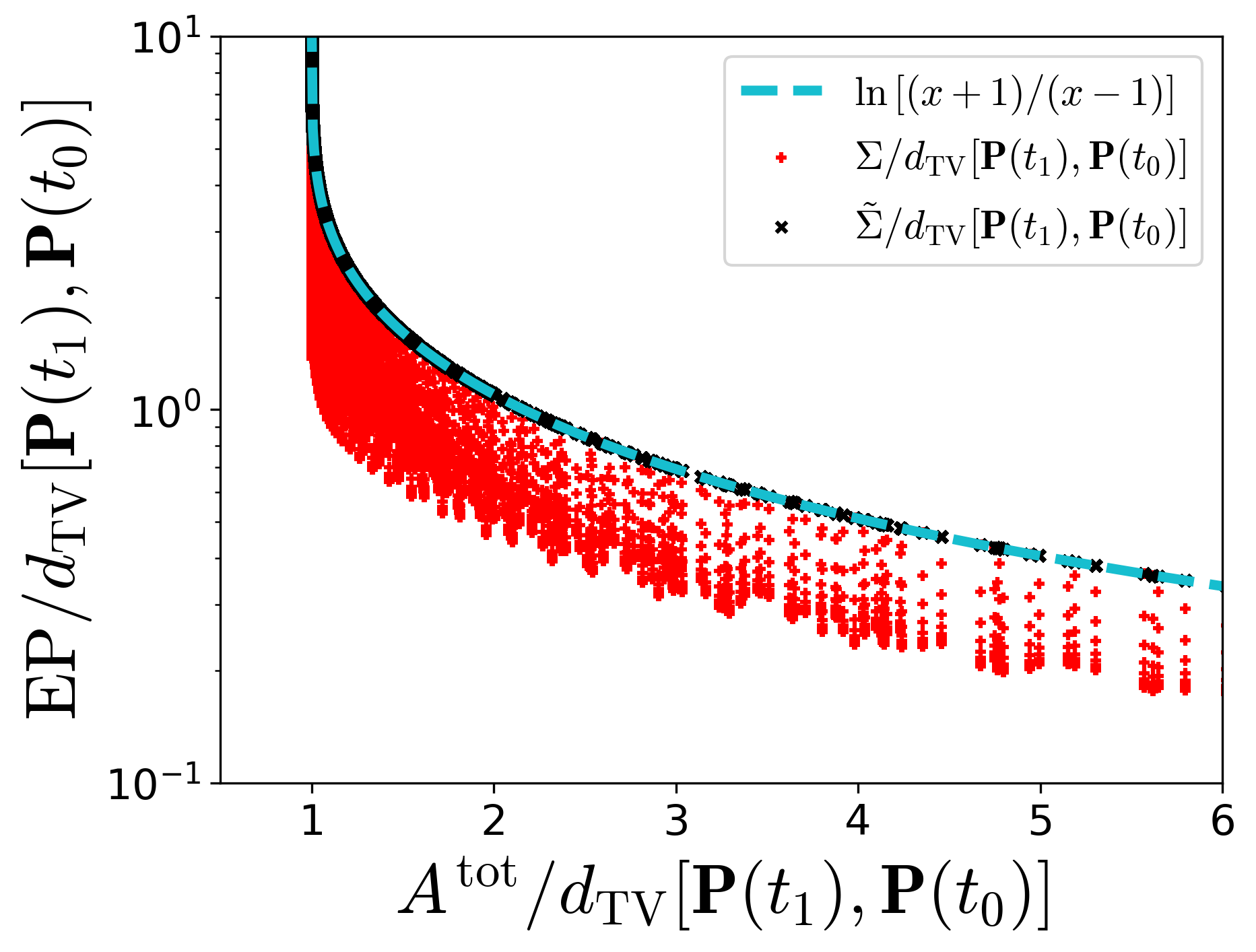}
\vskip -0.1in
\caption{
Plot of EP versus total dynamical activity for one-time-step relaxation process of the two-state system presented in Sec.~\ref{subsec:example1}. Both the $x$ and $y$ axes are scaled by the total variation distance between the initial and final state probabilities for each point. $+$ and $\times$ symbols represent scaled time-backward EP and time-reversed EP, respectively. The cyan dashed line denotes the speed limit bound estimated from $h(x)$ function. The parameters used for this plot are as follows: 
 $\mu = T = 1$, $ 0 \leq E_1\leq 10 $, $ 0 \leq \Delta t \leq 20 $ and $0.001 \leq g \leq 0.99$.
 }
\label{fig:onestep}
\vskip -0.1in
\end{figure}

Additionally, by applying Eqs.~\eqref{ineq:Jensen_dTV} and \eqref{ineq:triangle_dTV} to Eq.~\eqref{eq:stepwise_sl_uni}, we obtain the interval speed limit with unidirectional transitions as well, which is
\begin{equation}
    \tilde{\Sigma}^{({\rm b})} \geq A^{({\rm b})}_\textrm{tot} h \left( \frac{d_\textrm{TV} [\mathbf{P} (t_{\rm F}), \mathbf{P} (t_{\rm I})] - A^{({\rm u})}_\textrm{tot}}{A^{({\rm b})}_\textrm{tot}} \right)
\label{eq:unispeedlimitfirstform}
\end{equation}
with 
$A^{(\alpha)}_\textrm{tot} = \sum_{i=0}^{N-1} A^{(\alpha)} (t_i)$.
Similarly, Eq.~\eqref{eq:unispeedlimitfirstform} is valid when the argument of $h(x)$ is positive.
Thus, if $A^{({\rm u})} (t_i)$ is too large, {Eq.~\eqref{eq:unispeedlimitfirstform} can be violated.
Equations~\eqref{eq:stepwise_sl_uni} and \eqref{eq:unispeedlimitfirstform}  clearly show that the bound is modified by the dynamical activities associated with both types of transitions.
Since the contribution of unidirectional transitions is not included in $\tilde{\Sigma}^{({\rm b})}$, the dynamical activity associated with unidirectional transitions is subtracted from the total variation distance, which tends to decrease the bound of $\tilde{\Sigma}^{\rm(b)}$.

We note that there is a speed limit for 
a continuous-time process with unidirectional transitions in terms of a thermodynamic metric~\cite{Gupta2020Tighter}, which differs from the time-continuum limit of \eqref{eq:unispeedlimitfirstform}.

\section{Numerical verification}
\label{sec:Numerics}

In this section, we numerically verify our results through three toy models: the relaxation process of a two-state system, a cyclic process of a two-state system, and a relaxation process involving a unidirectional transition.
Additionally, we apply our results to cell-state dynamics of human breast cancer and deterministic finite automaton for identifying the multiples of 4.

\subsection{Relaxation process of a two-level system}\label{subsec:example1}

\begin{figure}[t!]
\includegraphics[width=0.48\textwidth]{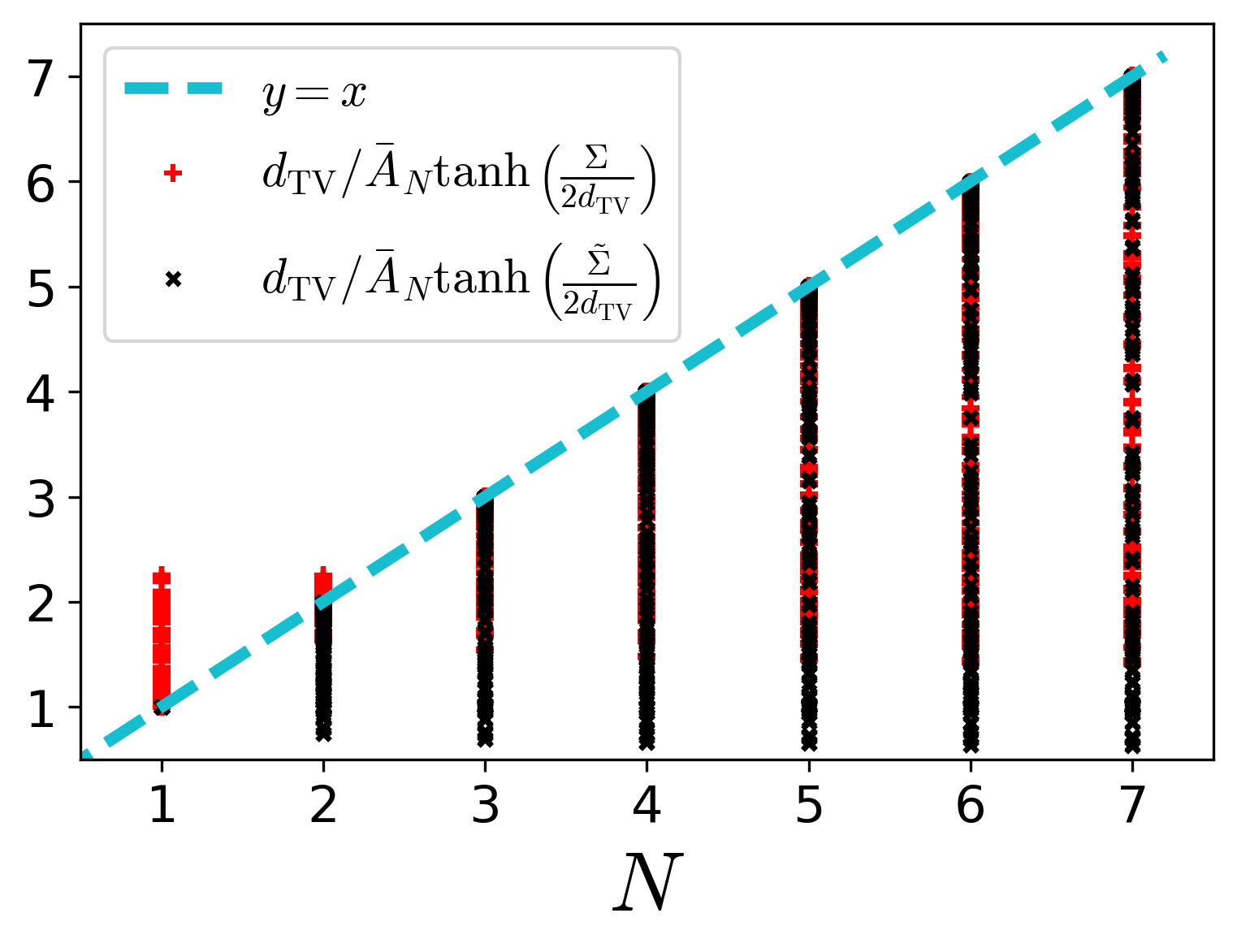}
\vskip -0.1in
\caption{
Plot of the right-hand side of Eq.~\eqref{eq:step_bound} versus the number of steps, denoted as $N$, for the $N$-step relaxation process of the two-state system presented in Sec.~\ref{subsec:example1}. $+$ and $\times$ symbols represent the bound on the number of steps in terms of the time-backward EP without the corrections term, $\sum_{i=0}^{N-1} 
    \Delta \Sigma'(t_i)$ and the time-reversed EP, respectively. The cyan dashed line denotes $y = x $ function. The parameters used for this plot are the same as those in the previous figure. 
}
\label{fig:Nstep_bound}
\vskip -0.1in
\end{figure} 

The first example considers a relaxation process with two states, labeled as $0$ and $1$. The energy gap between these  states is denoted as $E_1$. To determine the transition probabilities $M_{n,m}$ for a discrete-time process, we begin by examining a continuous-time relaxation process for the two-state system, which is in contact with a thermal reservoir at temperature $T$. The master equation of this continuous-time dynamics is 
\begin{equation} \label{eq:cont_time_master_ex1}
    \dot{\mathbf{P}} (t) = \mathbf{R} \cdot \mathbf{P} (t),
\end{equation}
where the transition rates $R_{1,0} = \mu \gamma$ and $R_{0,1} = \mu (1 - \gamma)$ follows the Arrhenius rule with $\gamma = e^{-E_1/T}/(1 + e^{-E_1/T})$. 
The diagonal elements of $\mathbf{R}$ are determined by $R_{0,0} = -R_{1,0}$ and $R_{1,1} = -R_{0,1}$ due to the normalization condition $\sum_n R_{n,m} = 0$. 
The system's probability vector is defined as $\mathbf{P}(t) = [P_0(t),P_1(t)]^{\textsf T}$.
The formal solution of Eq.~\eqref{eq:cont_time_master_ex1} is 
\begin{equation} \label{eq:ex1_discrete_eq}
    \mathbf{P}(t_1) = e^{\mathbf{R} (t_1 - t_0)} \cdot \mathbf{P}(t_0) \equiv \mathbf{M} \cdot \mathbf{P}(t_0),
\end{equation}
where the transition probability matrix is given by
\begin{equation}
    \mathbf{M} =
    \begin{pmatrix}
        1 - \gamma ( 1- e^{-\mu \Delta t}) &
        (1 - \gamma) ( 1- e^{-\mu \Delta t}) \\
        \gamma ( 1- e^{-\mu \Delta t}) &
        \gamma + (1 - \gamma) e^{-\mu \Delta t}
    \end{pmatrix}
\end{equation}
with the time interval $\Delta t = t_1 - t_0$. Equation~\eqref{eq:ex1_discrete_eq} serves as the equation of motion for our discrete-time process. 

Figure~\ref{fig:onestep} shows the plot of time-reversed and time-backward EPs with respect to the total dynamical activity. The data were generated from the one-step relaxation process of the two-state system. Note that the $x$ and $y$ axes of the plot are scaled by the  total variation distance for each point. The parameters $N = T = \mu = 1$ are used for this plot. The initial probabilities are parameterized as $P_0(t_0)= g$ and $P_1(t_0) = 1-g$. By choosing various values of $g$, $E_1$ and $\Delta t$, we evaluate the scaled EPs and $A_{\rm tot}$.  
The plot demonstrates that time-reversed EP saturates to the speed limit bound. This verifies the saturation condition discussed in Sec.~\ref{sec:saturation}.
On the other hand, the results for time-backward EP, represented by red cross symbols, are located below the bound, indicating that time-backward EP violates Eq.~\eqref{ineq:TwoPoint_SpeedLimit_rev_EP}.

To verify Eq.~\eqref{eq:step_bound}, we also calculate the lower bounds in $N$-step relaxation processes for $1\leq N\leq 7$.
Figure~\ref{fig:Nstep_bound} shows the results,
illustrating that for $N=1$, the time-reversed EP bound exactly matches $N$, in agreement with theoretical predictions.
For $N\leq2$, the time-backward EP bound, obtained by replacing $\tilde{\Sigma}$ with $\Sigma$, occasionally exceeds $N$,
indicating a violation of the conventional speed limit.
However, such violations become rare for $N>2$.
This observation arises because, in most processes, the time-reversed EP bound lies quite below $N$.
Only in slowly evolving processes, the time-reversed EP bound closely approaches $N$, but in this case $\Delta \Sigma'$ is too small to induce a violation.

\subsection{Cyclic steady-state process}
\label{sec:cyclicsteadystate}

\begin{figure}[t]
\includegraphics[width=0.48\textwidth]{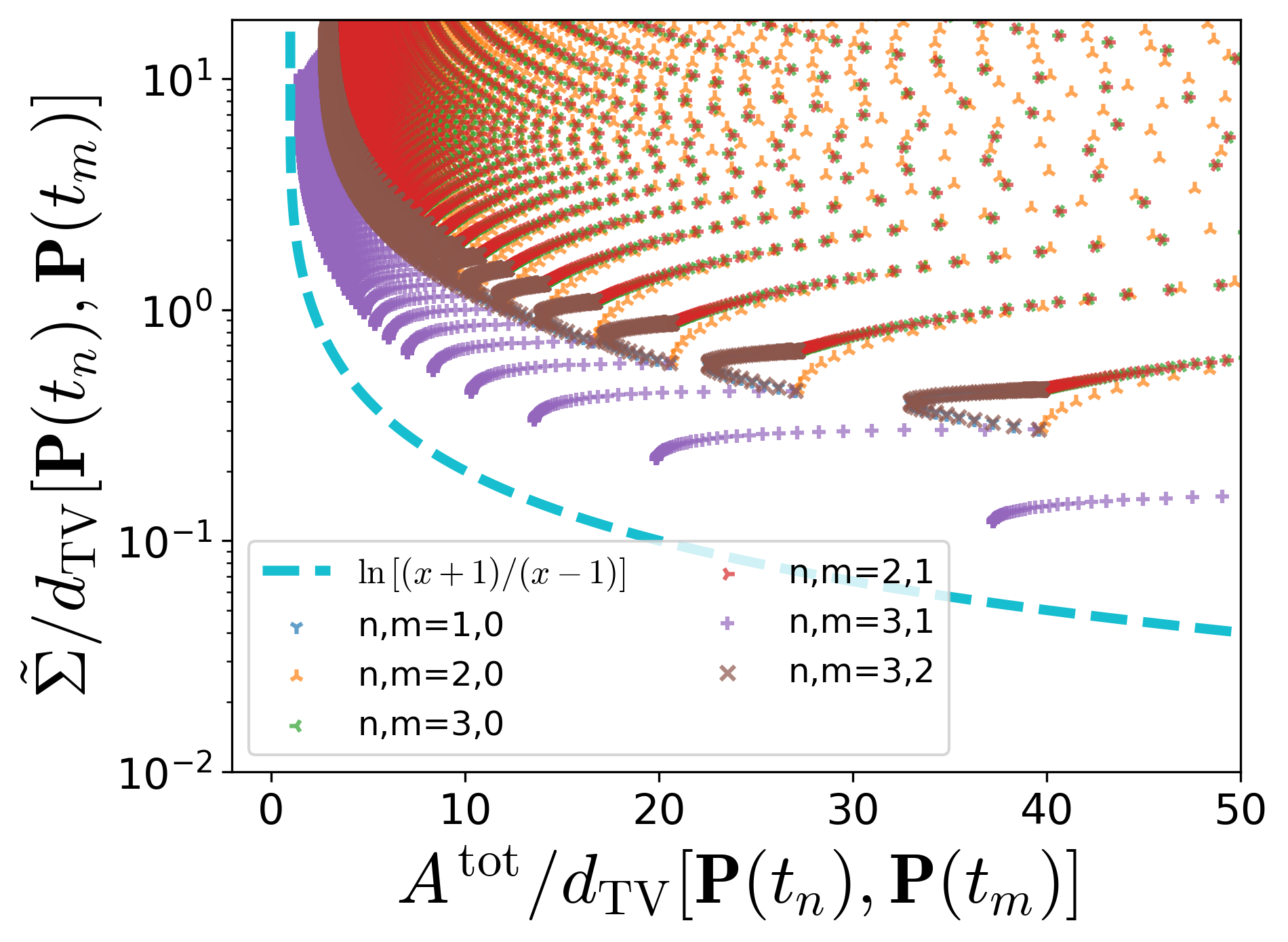}
\vskip -0.1in
\caption{ 
Plot of time-reversed EP  versus total dynamical activity for the cyclic steady-state process presented in Sec.~\ref{sec:cyclicsteadystate}. Both the $x$ and $y$ axes are scaled by the total variation distance between the initial and final state probabilities for each point. The dashed line denotes the interval speed limit. Different symbols represent different choices of $n$ and $m$ for total variation distance $d_{\rm TV}[\mathbf P (t_n),\mathbf P(t_m)] $.
The parameters used for this plot are as follows: $\mu = T = 1$, $E_0 = 0$, $0.01\leq \Delta E \leq 7 $, and $0.01\leq \Delta \tau \leq 7 $.
}
\label{fig:cyclictransition}
\vskip -0.1in
\end{figure}

The second example is a two-state system under periodic driving over time. The time period is $\tau = 4\Delta t$, that is, a four-step process. The equation of motion for the system is the same as Eq.~\eqref{eq:ex1_discrete_eq}, with time-dependent energy gap $E_1 (t)$: $E_1 (t_0) = \Delta E$, $E_1 (t_1) = 0$, $E_1 (t_2) = -\Delta E$, and $E_1 (t_3) = 0$. 

To verify the interval speed limit, we calculate the time-reversed EP and dynamical activity for the complete cycle, while also evaluating the total variation distance for all possible combinations of intermediate distributions at cyclic steady state. For this calculation, we set $\mu = T = 1$ and used various values of $\Delta E$ and $\Delta \tau$ within the range from $0.01$ to $7$. 
The calculation results are presented in Fig.~\ref{fig:cyclictransition}. The figure illustrates that for all non-zero total variation distances, the interval speed limit  provides meaningful bounds on EP. Especially, $d_\textrm{TV} [\mathbf{P} (t_3), \mathbf{P} (t_1)]$ yields the result closest to the bound.
This contrasts with the conventional speed limit, determined by the initial and final distributions, which yields a meaningless bound for a cyclic process. Thus, the results demonstrate the usefulness of the interval speed limit in EP estimation problem for a cyclic process.

\begin{figure}
\includegraphics[width=0.48\textwidth]{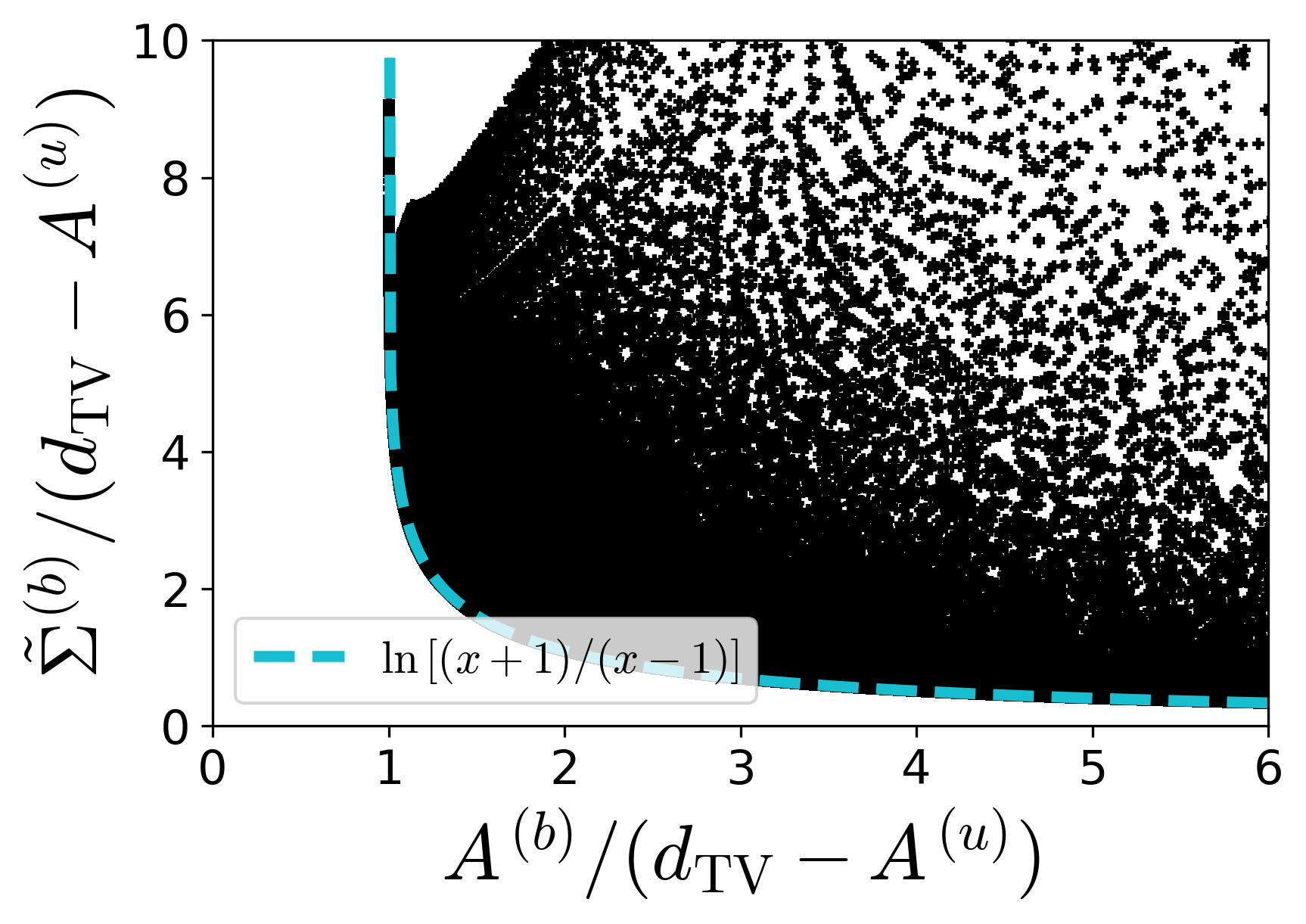}
\vskip -0.1in
\caption{ 
Plot of time-reversed EP versus dynamical activity of bidirectional transitions for the one-time-step relaxation process of the system, including a unidirectional transition, as presented in Sec.~\ref{sec:uni}. Both the $x$ and $y$ axes are scaled by the total variation distance minus unidirectional activity, $d_{\rm TV}[\mathbf{P}(t_1),\mathbf{P}(t_0)]-A^{({\rm u})}(t_0)$, for each point.
The dashed line represents the speed limit for a system including unidirectional transitions, given by Eq.~\eqref{eq:unispeedlimitfirstform}.  The black dots denote calculation results obtained from various values of $g$, $M_{1,0}$, and $M_{0,1}$ within the range ($0.01,0.9$) under the constraint $d_\textrm{TV}[\mathbf{P}(t_1),\mathbf{P}(t_0)] - A^{(u)}(t_0) > 0$. 
}
\label{fig:unitransition}
\vskip -0.1in
\end{figure}

\subsection{Three-level system with unidirectional transition}
\label{sec:uni}

The last example is a three-state model with one absorbing state indexed as $2$. The equation of motion is given by
\begin{equation}
    \mathbf{P}(t_{i+1}) = \left( \mathbf{M}^{({\rm b})} + \mathbf{M}^{({\rm u})} \right) \cdot \mathbf{P} (t_i),
\label{eq:uniMarkov}
\end{equation}
where the transition probability matrix for bidirectional transitions is 
\begin{align}
    \mathbf{M}^{({\rm b})} &=
    \begin{pmatrix}
        1 - M_{10} & M_{01} & 0 \\
        M_{10} & 1 - M_{01} - M_{21} & 0 \\
        0 & 0 & 1
    \end{pmatrix}
\end{align}
and the transition probability matrix for a unidirectional transition is 
\begin{align}
    \mathbf{M}^{(u)} &=
    \begin{pmatrix}
        0 & 0 & 0 \\
        0 & 0 & 0 \\
        0 & M_{21} & 0
    \end{pmatrix}.
\end{align} 
The initial condition is parameterized as $\mathbf{P} (t_0) = (g,1-g,0)^\mathrm{T}$, where $g$ denotes the initial probability of being in state $0$. Because there is no outgoing transitions from state $2$, the system eventually relaxes to the absorbing state with $P_n(\infty) = \delta_{n,2}$.

We evaluate time-reversed EP and dynamical activity for the one-time-step relaxation process described by Eq.~\eqref{eq:uniMarkov}. The results are plotted in Fig.~\ref{fig:unitransition}. The dashed line represents the bound of the speed limit for a system including unidirectional transitions, as given by Eq.~\eqref{eq:unispeedlimitfirstform}. Data points indicate the results obtained from various values of $g$, $M_{1,0}$, and $M_{0,1}$ chosen within the range $(0.01,0.9)$ under the constraint $d_\textrm{TV}[\mathbf{P}(t_1),\mathbf{P}(t_0)] - A^{(u)}(t_0) > 0$. This example validates the speed limit for a system consisting of unidirectional transitions and demonstrates its tightness.

\subsection{Cell-state dynamics and theoretical computation}
\label{sec:cancerAndcomp}

Finally, we consider two practical examples: cell-state dynamics~\cite{gupta2011stochastic} and a deterministic finite automaton~\cite{Gonzalo2024Thermodynamics}.
In Ref.~\cite{gupta2011stochastic}, the dynamics of phenotypic states in human breast cancer cells were investigated experimentally. The study focused on three distinct differentiation states of cancer cells: stem-like, basal, and luminal. Observations revealed that the distribution of these cell states evolved over time during culture growth. To explain the observed dynamics, a discrete-time Markov chain model was employed.

We apply our results to the model for a human breast cancer cell line, known as SUM159, in Ref~\cite{gupta2011stochastic}, where the transition matrix is given by
\begin{align}
  &  \mathbf{M}^{\rm (159)} =
\begin{pmatrix}
0.58 & 0.01 & 0.04 \\
0.35 & 0.99 & 0.49 \\
0.07 & 0.00 & 0.47  \\
\end{pmatrix} .
\label{eq:159_tranmat}\end{align}
Here the state distribution $P_n(t_i)$ for $n = 0,1, 2$ represents the proportions of stem-like, basal, and luminal states, receptively, after $i$ days of growth in culture.
The matrix for SUM 159 includes one unidirectional transition from the luminal state to the basal state.
In Fig.~\ref{fig:prac_example}, we plot the EP and the activity of bidirectional transitions for the one-step process of the breast cancer dynamics. The red `+' symbols represent the data. The results are well bounded by our inequality.

In addition, we validate our speed limits on a finite automaton, a type of computational machine, designed to identify multiples of 4 introduced in Ref.~\cite{Gonzalo2024Thermodynamics}.
The computational machine has four states, $n=0,1,2,3$, and initially starts from $n=0$.
Then, a given integer under identification is written in a string of binary numbers, either $0$ or $1$, which is taken sequentially as an input.
Depending on its current state and the input bit, the machine deterministically changes its state, which is designed to ensure that the final state is $0$ if and only if the given integer is a multiple of 4.
To investigate thermodynamics in this deterministic finite automaton, the authors considered a random input composed of independent and identically distributed random bits taking $0$ with probability $p_0$ and $1$ with probability $p_1$.
Then, the transition probability of the automaton is given by 
\begin{align}
    \mathbf{M}^{\rm (aut)} =
\begin{pmatrix}
p_0 & 0 & p_0 & 0 \\
p_1 & 0 & p_1 & 0 \\
0 & p_0 & 0 & p_0 \\
0 & p_1 & 0 & p_1
\end{pmatrix}.
\label{eq:automata_tranmat}
\end{align}
In Fig.~\ref{fig:prac_example}, black circle symbols depict the numerical results of the automaton calculated in the one-step process with random initial conditions.
Compared to the cell-state dynamics, the numerical data from the automaton dynamics spans a wider range of the plot.

\begin{figure}
\includegraphics[width=0.48\textwidth]{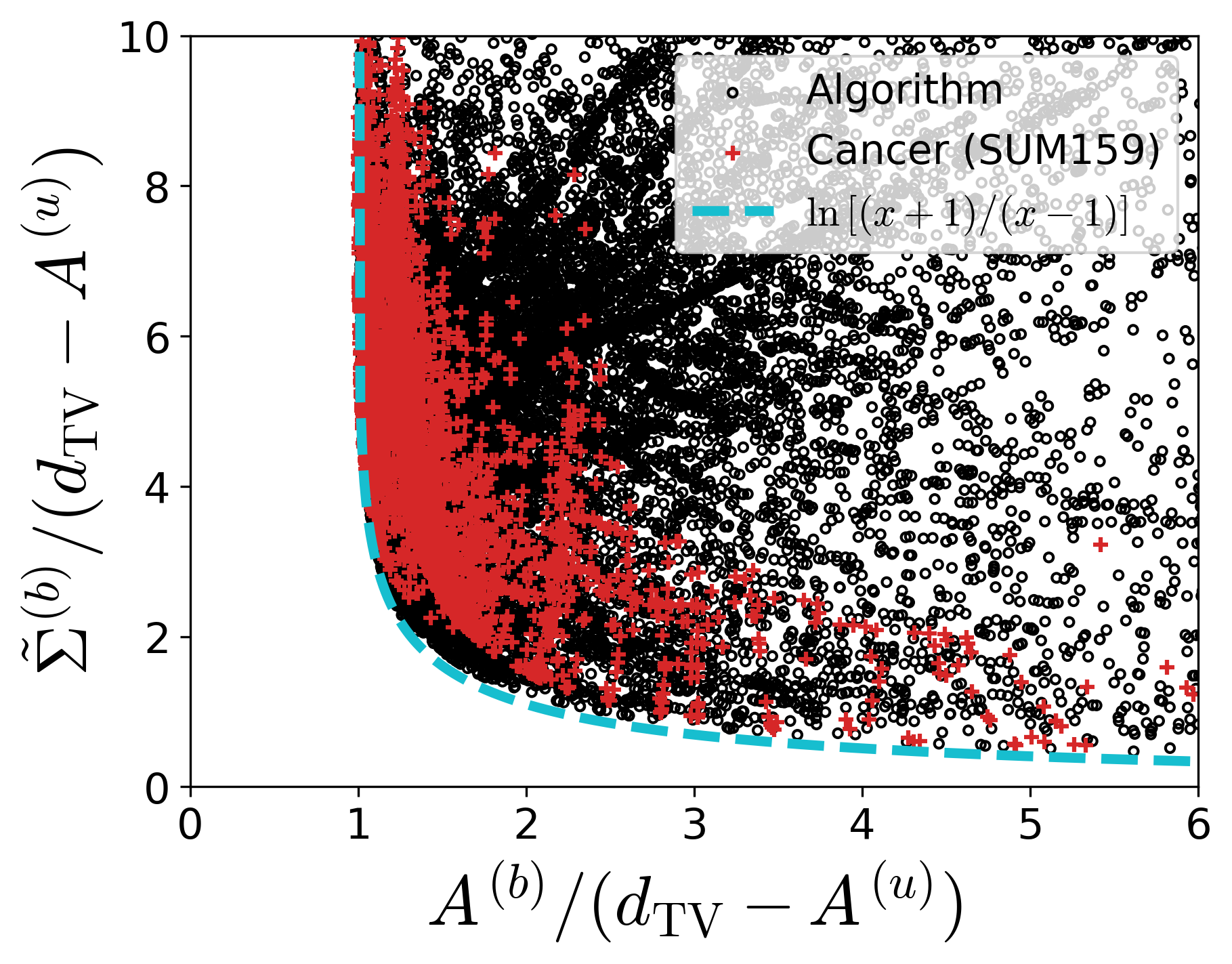}
\vskip -0.1in
\caption{ 
        Plot of the time-reversed EP versus the dynamical activity of bidirectional transitions for the one-step process of the breast cancer cell dynamics and the automaton dynamics in Sec.~\ref{sec:cancerAndcomp}. Both the $x$ and $y$ axes are scaled by the total variation distance minus unidirectional activity, $d_{\rm TV}[\mathbf{P}(t_1),\mathbf{P}(t_0)]-A^{({\rm u})}(t_0)$, for each point.
        The dashed line represents the speed limit for a system with unidirectional transitions, given by Eq.~\eqref{eq:unispeedlimitfirstform}. The black circles indicate the calculation results obtained from the automaton's dynamics by varying the initial condition, and the probabilities of its input, $p_0$ and $p_1$. The red `+' symbols represent the results from the breast cancer cell dynamics of SUM 159 under various initial conditions with the transition matrix given by Eq.~\eqref{eq:159_tranmat}.
}
\label{fig:prac_example}
\vskip -0.1in
\end{figure} 
\section{Conclusions}
\label{sec:disscussion}
We investigated the speed limits in discrete-time Markov chains. We found that the time-reversed EP satisfies the speed limit in the same form as the conventional relation, while time-backward EP violates this bound.
These findings highlight significant differences from previous extensions of TUR to discrete-time processes.
First, the conventional TUR is not valid for discrete-time processes, necessitating modifications such as the exponential TUR and the linear TUR~\cite{proesmans2017discrete, Liu2020Thermodynamics}. 
Second, both discrete-time TURs are not applicable to systems with time-dependent driving.
In particular, the exponential TUR holds only in the steady state and the linear TUR holds for relaxation processes without time-dependent driving.
Lastly, the bound of the exponential TUR rapidly decays as the system is driven further from equilibrium, and the linear TUR bound vanishes if at least one staying probability at any state is zero.
In contrast, the examples in Sec.~\ref{sec:Numerics} demonstrate that our results hold for general time-dependent processes even in the presence of zero staying probabilities, providing a meaningful bound.
Therefore, we anticipate that the discrete-time speed limit can offer a more efficient method for estimating EP in Markov chains compared to discrete-time TUR.

We also derived several variants of the conventional speed limit. The first is the stepwise speed limit which offers a bound in terms of distances between two consecutive probability distributions at all time steps. 
Second, we derived the interval speed limit  which is expressed solely by a single distance between any intermediate-time points. This relation is useful especially when only partial-time information on probability distributions is available. In addition, this relation can provide a nontrivial bound even in systems driven by a cyclic protocol and generally provides a tighter bound. 
Finally, we obtained a modified speed limit for systems with unidirectional transitions. This relation can be applied to various systems with absorbing states such as extinction dynamics and opinion dynamics.

Our derivation approaches are also applicable to continuous-time Markov processes, ensuring the validity of the derived relations in continuous-time processes. We believe that our findings will contribute to exploring thermodynamics in both discrete-time and continuous-time systems, where existing speed limits may not be applicable.

\begin{acknowledgments}
This research was supported by the KIAS Individual Grant Nos. PG081802 (S.L.), and PG064902 (J.S.L.) at the Korea Institute for Advanced Study and an appointment to the JRG Program at the APCTP through the Science and Technology Promotion Fund and Lottery Fund of the Korean Government (J.-M.P.). This was also supported by the Korean Local Governments - Gyeongsangbuk-do Province and Pohang City (J.-M.P.). This work was supported by the National Research Foundation of Korea(NRF) grant funded by the Korea government(MSIT) (RS-2025-00557038) (J.-M. P.).
S.L. acknowledges additional support from Deutsche
Forschungsgemeinschaft (DFG, German Research Foundation) within SFB 1551
- project number 464588647.
S.L. thanks Seung Ki Baek for valuable comments and references. S.L. also acknowledge the warm hospitality of Christopher Jarzynski at the Institute for Physical Science and Technology Department at the University of Maryland, College Park.
\end{acknowledgments}

\appendix
\renewcommand\thefigure{\thesection.\arabic{figure}}
\setcounter{figure}{0}

\section{Differences between $\Delta \Sigma$ and $\Delta \tilde{\Sigma}$}
\label{app:differences_EPs}

In this section, we compare $\Delta \Sigma$ and $\Delta \tilde{\Sigma}$ to describe their differences in several aspects.

\subsection{Contributions of staying events}
The transitions staying in the same state, occurring with probability $M_{n,n}(t_i)$, gives no contributions to the time-reversed EP $\Delta \tilde{\Sigma}$ as in the continuous-time EP rate. In fact, $\Delta \tilde{\Sigma}$ have the same mathematical form as the continuous-time EP rate, given by
\begin{align}
    \dot\Sigma_\textrm{cont} (t)
    = \sum_{n,m}
    R_{n,m} (t) P_m(t)
    \ln \frac{R_{n,m} (t) P_m(t)}{R_{m,n} (t) P_n(t)}.
\end{align}
In contrast, the staying events contribute to time-backward EP.

\subsection{Decomposition with system entropy change}
The system entropy, defined as the Shannon entropy, is given by
\begin{align}
    S_{\rm sys }(t_i) \equiv - \sum_n P_n(t_i)\ln P_n(t_i).
\end{align}
We can decompose the time-backward EP into
\begin{align}
    \Delta \Sigma (t_i) = \Delta S_{\rm sys} (t_i) + \Delta S_\textrm{b} (t_i),
\end{align}
where $\Delta S_\textrm{sys} = S_\textrm{sys}(t_{i+1}) - S_\textrm{sys}(t_{i})$ and
\begin{align}
     \Delta S_\textrm{b} (t_i) \equiv \sum_{n,m} M_{n,m} (t_i) P_m(t_i) \ln \frac{M_{n,m} (t_i) }{M_{m,n} (t_i) }.
\end{align}
This decomposition is reminiscent of the thermodynamic EP rate for continuous-time processes, which is given by
\begin{align}
    \dot\Sigma_\textrm{cont} (t)
    =& \dot S_\text{sys}(t) + \dot{S}_\textrm{env}(t)
\label{def:EP_cont}, \end{align}
where the environmental entropy production rate is defined as
\begin{align}
    \dot{S}_\textrm{env}(t) = \sum_{n,m}
    R_{n,m} (t) P_m(t)
    \ln \frac{R_{n,m} (t)}{R_{m,n} (t)}.
\end{align}
For this reason, $\Delta S_\textrm{b}$ is referred to as the EP of the heat reservoirs in discrete-time process in Ref.~\cite{Liu2020Thermodynamics}.

For the time-backward EP,  the decomposition is given by
\begin{align}
    \Delta \tilde{\Sigma} (t_i)= \Delta S_{\rm sys} +  \Delta \tilde{S}_\textrm{b}.
\end{align}
with the entropy flow~\cite{gaspard2004time}
\begin{align}
    \Delta \tilde{S}_\textrm{b} (t_i) \equiv \sum_{n,m} M_{n,m} (t_i) P_m(t_i) \ln \frac{M_{n,m} (t_i) P_n(t_{i+1})}{ M_{m,n} (t_i) P_n(t_i) }.
\end{align}
Unlike $\Delta S_\textrm{b}$ and the environmental EP in continuous-time processes, the summand in $\Delta \tilde{S}_\textrm{b}$ depends on $P_n(t)$.

\subsection{Relation with detailed balance condition}
Since time-reversed EP takes the form of Kullback-Leibler divergence, it vanishes if and only if $M_{n,m}(t_i) P_m(t_i) = M_{m,n}(t_i) P_n(t_{i})$, which is called detailed balance condition in Ref.~\cite{ge2006reversibility}.
Since detailed balance ensures that the probability distribution remains unchanged, $P_n(t_{i+1}) = P_n(t_i)$,
Eq.~\eqref{eq:relationbtwtimeandback} shows that the time-backward EP also vanishes when the detailed balance is satisfied.

However, the detailed balance is not a necessary condition for vanishing time-backward EP.
One simple example is a single bit under the NOT gate operation.
Here, the bit system has two states $n=0,1$ and the transition probability matrix is given by
\begin{align}
    \mathbf{M}=
    \begin{pmatrix}
        0 & 1 \\ 1 & 0
    \end{pmatrix}.
\end{align}
When the system starts from the initial condition $P_n(t_0) = \delta_{n,0}$, the time-backward EP takes zero although the detailed balance condition is not obeyed in this process.
However, the time-reversed EP is non-zero and even diverges.
This example clearly demonstrates that the time-backward EP is not a proper indicator of the detailed balance condition.

The explicit condition for vanishing time-backward EP is $M_{n,m}(t_i) P_m(t_i) = M_{m,n}(t_i) P_n(t_{i+1})$.
By applying Bayes’ theorem, this condition can be rewritten as
$M_{m,n}(t_i) = M_{n,m}(t_i) P_m(t_i)/P_n(t_{i+1})= P(m,t_i|n,t_{i+1})$ with the conditional probability $P(m,t_i|n,t_{i+1})$ that the system was in the state $m$ at time $t_i$ given that it is in $n$ at time $t_{i+1}$.
This condition reflects that the backward time evolution matches the forward evolution, $P(n,t_{i+1}|m,t_i) = P(m,t_i|n,t_{i+1})$, and it can hold even when detailed balance is violated as shown in the NOT-gate example.

\subsection{Relation with the H-theorem}

In relaxation processes, where the transition probabilities are time-independent, the time-backward EP relates to the $H$-function, a monotonically decreasing quantity in time, defined as~\cite{morimoto1963markov}
\begin{align}
H(t_i) \equiv \sum_n P_n (t_i) \ln \frac{P_n(t_i)}{P_n^{\rm eq}}
\end{align}
where $P_n^{\rm eq}$ is the equilibrium distribution.
As shown in~\cite{Lee2018Derivation}, the time-backward EP can be expressed as
\begin{align}
\Delta \Sigma(t_i) = H(t_i) - H(t_{i+1})\geq0,
\end{align}
indicating the connection between the time-backward EP and the $H$-theorem.

We note that it is a special case of the intrinsic mismatch cost, introduced in Ref.~\cite{kolchinsky2017dependence,Gonzalo2024Thermodynamics}, defined as
\begin{align}
    \Delta \Sigma_\textrm{miss} (t_i)
    = \sum_n P_n (t_i) \ln \frac{P_n (t_i)}{r_n (t_i)}
    - \sum_n P_n (t_{i+1}) \ln \frac{P_n (t_{i+1})}{r_n (t_{i+1})}
\end{align}
with an arbitrary reference probability $r_n(t_i)$ and $r_n(t_{i+1})=\sum_m M_{n,m}(t_i) r_n(t_i)$.
If we choose $r_n(t_i) = P_n^\textrm{eq}$, then it reduce to the time-backward EP.

\section{Derivation of Eq.~(10)}
\label{app:derivation_symGilardoni}
In this section, we revisit Ref.~\cite{falasco2022beyond} to represent the details for the derivation of the inequality in~\eqref{ineq:Jeffreys_vs_dTV}.
To do this, we first derive two auxiliary relations. The first relation is an ordering between the total variation distance $d_\textrm{TV} [\mathbf{p}, \mathbf{q}]$ in Eq.~\eqref{def:TVD} and the Le Cam's distance defined as
\begin{equation}
    d_\textrm{LeCam} [\mathbf{p}, \mathbf{q}]
    \equiv \sqrt{\frac{1}{2} \sum_n \frac{(p_n - q_n)^2}{p_n + q_n}}.
\end{equation}
By using the Cauchy-Schwarz inequality, we can show that
\begin{align}
    \left( 2 d_\textrm{TV} [\mathbf{p}, \mathbf{q}] \right)^2
    &= \left( \sum_n |p_n - q_n| \right)^2 \nonumber \\
    &= \left( \sum_n \frac{|p_n - q_n|}{\sqrt{p_n + q_n}} \sqrt{p_n + q_n} \right)^2 \nonumber \\
    &\leq \left( \sum_n \frac{(p_n - q_n)^2}{p_n + q_n} \right)
    \left( \sum_n (p_n + q_n) \right) \nonumber \\
    &= \left ( 2 d_\textrm{LeCam} [\mathbf{p}, \mathbf{q}] \right)^2.
\end{align}
Since both distances are positive, it yields $d_\textrm{TV} [\mathbf{p}, \mathbf{q}] \leq d_\textrm{LeCam} [\mathbf{p}, \mathbf{q}]$.

To find the second relation, we express the squared Le Cam's distance as
\begin{align}
    \left( d_\textrm{LeCam} [\mathbf{p}, \mathbf{q}] \right)^2
    = \frac{1}{2} \sum_n |p_n - q_n| \tanh \left( \frac{1}{2} \left| \ln \frac{p_n}{q_n} \right|\right).
\end{align}
Here we used the property that $p_n - q_n$ and $\ln(p_n/q_n)$ have the same sign.
Since $\tanh x$ is concave for $x\geq0$, by introducing a probability $r_n \equiv |p_n - q_n|/(2d_\textrm{TV}[\mathbf{p},\mathbf{q}]) \geq 0$ and applying the Jensen's inequality, we find
\begin{align}
    \left( d_\textrm{LeCam} [\mathbf{p}, \mathbf{q}] \right)^2 &= d_\textrm{TV}[\mathbf{p},\mathbf{q}]
    \sum_n r_n \tanh \left( \frac{1}{2} \left| \ln \frac{p_n}{q_n} \right| \right) \nonumber \\
    &\leq d_\textrm{TV}[\mathbf{p},\mathbf{q}] \tanh \left( \frac{1}{2} \sum_nr_n \left| \ln \frac{p_n}{q_n} \right| \right) \nonumber \\
    &= d_\textrm{TV}[\mathbf{p},\mathbf{q}] \tanh \left( \frac{\sum_n (p_n - q_n) \ln \frac{p_n}{q_n}}{4 d_\textrm{TV}[\mathbf{p},\mathbf{q}]} \right)\nonumber \\
    &= d_\textrm{TV}[\mathbf{p},\mathbf{q}] \tanh \left( \frac{D [ \mathbf{p} \parallel \mathbf{q}] + D [ \mathbf{q} \parallel \mathbf{p}]}{4 d_\textrm{TV}[\mathbf{p},\mathbf{q}]}  \right).
\end{align}

Consequently, we arrive at
\begin{align}
    (d_\textrm{TV}[\mathbf{p},\mathbf{q}])^2
    &\leq (d_\textrm{LeCam}[\mathbf{p},\mathbf{q}])^2 \nonumber \\
    &\leq d_\textrm{TV}[\mathbf{p},\mathbf{q}] \tanh \left( \frac{D_{\rm J} [\mathbf{p}, \mathbf{q}]}{2 d_\textrm{TV}[\mathbf{p},\mathbf{q}]}  \right).
\end{align}
As $\tanh^{-1}x = (1/2) \ln ((1+x)/(1-x))$, by rearranging this inequality, we finally obtain
\begin{align}
    d_\textrm{TV}[\mathbf{p},\mathbf{q}]
    \ln \frac{1 + d_\textrm{TV}[\mathbf{p},\mathbf{q}]}{1 -d_\textrm{TV}[\mathbf{p},\mathbf{q}]}
    \leq D_{\rm J}[\mathbf{p},\mathbf{q}],
\end{align}
which is equivalent to Eq.~\eqref{ineq:Jeffreys_vs_dTV}.

\bibliography{discrete}

\end{document}